\documentclass[aps,twocolumn]{revtex4}
\usepackage{graphicx,subfigure}
\usepackage{amsmath,amssymb}
\usepackage{bm}
\newcommand{\Dslash}{D\hspace{-1.6ex}/\hspace{0.6ex} }
\newcommand{\Wslash}{W\hspace{-1.6ex}/\hspace{0.6ex} }
\newcommand{\pslash}{p\hspace{-1.6ex}/\hspace{0.6ex} }
\newcommand{\partslash}{\partial\hspace{-1.6ex}/\hspace{0.6ex} }
\newcommand{\be}{\begin{eqnarray}}
\newcommand{\ee}{\end{eqnarray}}
\newcommand{\bear}{\begin{eqnarray}}
\newcommand{\eear}{\end{eqnarray}}
\newcommand{\ba}{\begin{array}}
\newcommand{\ea}{\end{array}}

\begin{document}

\title{CP violation during the electroweak sphaleron transitions}

\author{  Edward  Shuryak and Ismail Zahed}

\affiliation{Department of Physics and Astronomy, Stony Brook University,
Stony Brook NY 11794-3800, USA}

\begin{abstract}
We suggest a  specific semiclassical background field,  the so called pure gauge sphaleron explosion, 
to evaluate the magnitude of the CP violation stemming from the standard phase of the CKM matrix.
We use it to evaluate the matrix elements of some next-to-leading order effective CP-violating operators
suggested in the literature.  We also derive the scale dependence of the corresponding coefficients.  
Finally, we discuss the expected magnitude of the CP violation in the cold electroweak scenario.

\end{abstract}
\maketitle
\section{Introduction}
It is by now a standard statement, repeated at the beginning of (nearly) any talk on 
baryogenesis, that while the Standard Model (SM) includes nonzero effects for all three Sakharov's ingredients \cite{Sakharov:1967dj} -- baryon number violation, CP violation and deviation from equilibrium --
their products falls short of the observed baryonic asymmetry of the Universe
\be n_B/n_\gamma\sim 6*10^{-10}  \label{eqn_ba}\ee
by many orders of magnitude.  

As a result, the mainstream of baryogenesis studies focus mostly on  ``beyond the Standard Model" (BSM)
scenarios, in which  new sources of CP violation are introduced, e.g. in the extended Higgs or neutrino sector. Leptogenesis scenarios are based on superheavy neutrino decays, occurring at very high scales
and satisfying both large CP and out-of-equilibrium requirements, with lepton asymmetry
then transformed into baryon asymmetry at the electroweak scale.  
However, as all BSM scenarios remain at this time purely hypothetical, without support  from 
the LHC and other experiments so far, perhaps it is warranted to revisit the SM-based  scenarios.



While most of this paper will be focused on the  CP violation during baryon-number producing sphaleron transitions,
let us here comment on the third necessary ingredient, a deviation from thermal equilibrium.
Standard cosmology assumes that reheating and entropy production of the Universe take place at a
scale much higher than the electroweak scale. In addition,  the standard model 
with the Higgs mass at $125 \, GeV$ has an electroweak transition only of a smooth crossover type.
If the assumption is correct, and there would be no new particles at the electroweak scale found, 
the transition would be  very  smooth, without  significant out-of-equilibrium effects. 

A non-standard
cosmological scenario  \cite{GarciaBellido:1999sv,Krauss:1999ng}  known as ``cold electroweak baryogenesis"
assumed that the inflation era ends at the electroweak scale. 
The simplest explcit model studied  ties two scalars, the $inflaton$ and the Higgs. The word
``Cold" used in the name refers to  the fact that 
at the end of the reheating and equilibration of the Universe, the temperature is  $T=30- 40\, GeV$, well  below the
critical electroweak temperature $T_c$.
What this does is to ensure that  one of Sakharov's conditions -- out of equilibrium --
 is  satisfied $maximally$. It also suggests that we should not worry  about wash-outs of any asymmetry developed earlier.


 Extensive numerical simulations \cite{GarciaBellido:1999sv,Smit_etal} had found 
 several phenomena, not anticipated before. One of them is the existence of metastable
{\em  hot spots}, relatively long-lived domains of the symmetric (unbroken) phase with a small
Higgs vacuum expectation value (VEV) and  large  magnitude  gauge fields within, balancing the pressure.
The other was a rather high rate of the sphaleron-like baryon number violating
transitions, all occurring inside those  hot spots.

Metastability of  hot spots was later explained when they were identified with the multi-quanta 
bags \cite{Crichigno:2010ky}. The numerically calculated sphaleron rates were related to the
analytic solution, the so called COS sphalerons in \cite{Flambaum:2010fp}. The key point was that
 the scale of the gauge fields -- and thus the height of the barrier to climb --
 is not determined by the Higgs VEV in the broken phase, but by the size of the
  hot spot $\rho$ with the unbroken phase. The larger the spot, the lower the barrier to climb,
  and the larger are the barrier penetration rates.

 This scenario was further developed in \cite{Flambaum:2010fp}
  by acknowledging a very significant role of the 
 top (antitop)  quarks,  not included in numerical simulations for technical reasons. 
 The sphaleron rate is further increased if standard top production in sphaleron transition
  (usually assumed) is substituted by the process in which  the pre-existing anti-tops in the bag are ``eaten up".
 
 CP violation in this scenario was studied in \cite{Burnier:2011wx}, where it was shown 
that the 4-th order weak quark decay diagrams can  create  an asymmetry in the quark/antiquark diffusion rates
via the known CP-violating phase of the CKM matrix.   
 The  estimates derived  in  \cite{Burnier:2011wx} lead to  an asymmetry in the top/antitop population  
 within the  bags of the order
 of $10^{-10}$. It is close to what is needed for the explanation of the baryon asymmetry, but not enough.
 The main question discussed in this paper 
 is that  CP violation may take place  $simultaneously$ with
 the process of baryon number violation, during the
 sphaleron production and explosion.  
%

\section{The exploding pure-gauge sphaleron} 

Both static and time-dependent exploding solutions for the pure-gauge sphaleron have been originally discussed 
by Carter,Ostrovsky and Shuryak (COS)  \cite{Ostrovsky:2002cg}. Its simpler derivation, to be used below, has been subsequently found by us in 
\cite{Shuryak:2002qz}.
The technique relies on an  {\em off-center conformal transformation} of the $O(4)$ symmetric Euclidean instanton
solution, which is analytically continued to  Minkowski space-time. The focus of that work
\cite{Shuryak:2002qz} was however on  the detailed description of the fermion production. 

The original $O(4)$-symmetric solution is given by the following ansatz
\be  g A_\mu^a=\eta_{a\mu\nu} \partial_\nu F(y), \,\,\,\, F(y)=2\int_0^{\xi(y)} d\xi'   f(\xi')     \ee
with $\xi=  ln(y^2/\rho^2)$ and $\eta$ the 't Hooft symbol. 
Upon substitution of the gauge fields in  the gauge Lagrangian 
one finds  the effective  action for $f(\xi)$ 
\be S_{eff}=   \int d\xi \left[{\dot{f}^2\over 2}+2f^2(1-f)^2 \right]
\ee   
corresponding to the motion of a particle in a double-well potential. 
In the Euclidean formulation, as written, the effective potential is inverted
\be V_E=-2f^2(1-f)^2 \ee
and the corresponding solution  is the well known BPST instanton, 
a path connecting the two maxima of $V_E$, at $f=0, f=1$. 
Any other solution of the  equation of motion
following from $S_{eff}$
obviously generalizes to a solution of the Yang-Mills equations for $A_\mu^a(x)$ as well.

The next step is to perform an off-center conformal transformation 
\be (x+a)_\mu={2 \rho^2 \over (y+a)^2} (y+a)_\mu
\ee
with $a_\mu=(0,0, 0, \rho) $.  It changes the original spherically symmetric 
solution 
to a solution of Yang-Mills equation depending on new coordinates
 $x_\mu$, with separate dependences on time
$x_4$ and the 3-dimensional radius $r=\sqrt{x_1^2+x_2^2+x_3^2}$. 

The last step  is the analytic continuation to Minkowski time $t$, via $x_4\rightarrow i t$. 
The original parameter $\xi$ in terms of 
these   Minkowskian coordinates, which we still call  $x_\mu$, has the form
\be \xi ={1\over 2}  log{y^2\over \rho^2}={1\over 2} log\left( {(t+i\rho) ^2-r^2 \over (t-i\rho) ^2-r^2 } \right) \ee
which is pure imaginary.To avoid carrying the extra $i$, we use the real 
\be \xi_E \rightarrow -i \xi_M = arctan\left( { 2 \rho t \over t^2-r^2-\rho^2 } \right)   \label{arctan}  \ee 
and in what follows we will drop the suffix $E$.
Switching from imaginary to real $\xi$,  correponds to switching from the Euclidean to
 Minkowski spacetime solution. It changes the sign of the acceleration, or the sign of the effective potential $V_M=-V_E$,
to that of the normal double-well problem. The static sphaleron solution corresponds
to a particle sitting on the potential maximum at $f=1/2$, and the sphaleron decay
to ``tumbling" paths. 

The needed solution of the equation of motion has been given in  \cite{Shuryak:2002qz} \footnote{There was a misprint in the index of this expression in the original paper.} 

\be f(\xi)={1 \over 2} \left[ 1- \sqrt{1+\sqrt{2\epsilon}} dn \left(  \sqrt{1+\sqrt{2\epsilon}} (\xi-K), {1 \over \sqrt{m}} \right) \right] 
\ee
where  $dn(z,k) $ is one of the elliptic Jacobi functions, $2\epsilon=E/E_s,2m=1+1/\sqrt{2\epsilon}$,
$E=V(f_{in})$ is the conserved energy of the mechanical system normalized to that of the sphaleron $E_s=V(f=1/2)=1/8$. 
 Since the start from exactly
 the maximum takes a divergent time, we will start  nearby the turning point
 with 
 \be f(0)=f_{in}={1\over 2} - \kappa, \,\,\,\,\,\, f'(0)=0 \ee
where a small displacement $\kappa$ ensures that ``rolling downhill" from the maximum takes a finite time and
that the half-period $K$ -- given by an elliptic integral -- in the expression is not divergent. 
In the plots below we will use $\kappa=0.01$, but the results  dependent on its value very weakly.

\begin{figure}[h]
\begin{center}
\includegraphics[width=8cm]{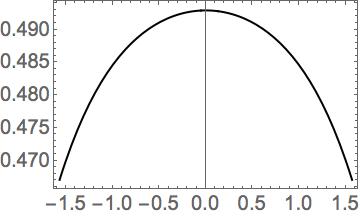}
\caption{The function $f(\xi)$ in the needed  range of its argument $\xi \in [-\pi/2,\pi/2]$}
\label{fig_f[ksi]}
\end{center}
\end{figure}

The solution above describes a particle tumbling periodically between two turning points, 
and so the expression above defines a periodic function for all $\xi$. However, as 
it is clear from (\ref{arctan}), for our particular application the only relevant domain is $\xi \in [-\pi/2,\pi/2]$. The solution $ f(\xi)$ in it 
is shown in Fig.~\ref{fig_f[ksi]} . Using the first 3 nonzero terms of its Taylor expansion
\be f\approx 0.49292875 - 0.0070691232 \xi^2 - 0.0011773\xi^4  \ee
$$- 0.0000781531899\xi^6 $$
we find a parametrization with an accuracy of $10^{-5}$, obviously invisible in the plot and
 more than enough for our considerations. 
 
The gauge potential has the form \cite{Shuryak:2002qz} 
\be gA_4^a=-f(\xi) { 8 t\rho x_a \over [(t-i\rho)^2-r^2]  [(t+i\rho)^2-r^2]  } \label{eqn_field} \ee
$$ gA^a_i=4\rho f(\xi) { \delta_{ai}(t^2-r^2+\rho^2)+2\rho \epsilon_{aij} x_j +2 x_i x_a \over [(t-i\rho)^2-r^2]  [(t+i\rho)^2-r^2]  } $$
which is manifestly real.  
From those potentials we
 generate rather lengthy expressions for the electric and magnetic fields,  and eventually for
 CP-violating operators, using Mathematica. We will not present those expressions in the paper,
 illustrating only some relevant features  by the plots.
 
\begin{figure}[t]
\begin{center}
\includegraphics[width=8cm]{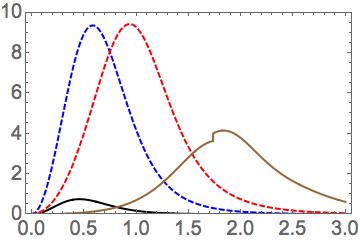}
\includegraphics[width=8cm]{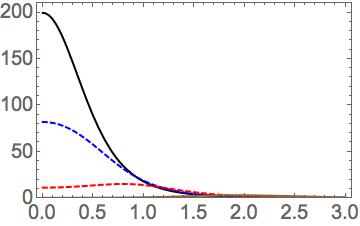}
\includegraphics[width=8cm]{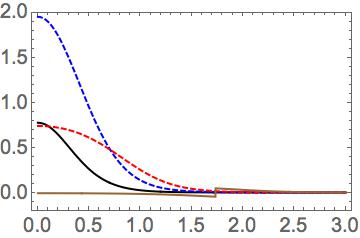}
\caption{Radial dependence of $\vec{E}^2$ (upper plot),   $\vec{B}^2$ (middle plot) and  $\vec E\cdot\vec B$
(lower plot). Different curves correspond to different time moments, as follows: solid black curve for $t=0.1$, dashed blue one for $t=0.5$,  red dashed one for $t=1 $ and brown solid one for $t=2$. 
}
\label{fig_fields}
\end{center}
\end{figure}

The electric field  squared at certain times is plotted versus distance in Fig.~\ref{fig_fields}.
Because we start at the ``turning point" with zero momentum, it
vanishes at $t=0$. As shown in Fig.~\ref{fig_fields} (upper), the electric field rapidly grows to some maximum,
and at late time it gets concentrated near a sphere expanding with the  speed of light. 
The magnetic field squared is shown in  Fig.~\ref{fig_fields} (middle). Note that
it starts with a COS sphaleron, and the initial values of the magnetic field is rather large.
The sphaleron is a ball with
3 ``colors"  of the field $\vec{B}^1,  \vec{B}^2,\vec{B}^3$ rotating around spatial axes 1,2,3, respectively,
but $(\vec{B}^a)^2$ is sphericaly symmetric. 
Asymptotically at large $t$ the magnetic field $\vec B$ becomes comparable in magnitude and normal to $\vec E$ and $\vec r$,
as expected  for a shell made up of weak fields made of  massless gauge bosons.

 The evolution of the Chern-Simons number is related to $\vec E\cdot \vec B$, which is proportional to the divergence of the
 topological Chern-Simons current. This quantity is displayed in Fig.~\ref{fig_fields} (lower). 
 Naively it is expected that the change of the Chern-Simons number 
  \be \Delta N_{CS} =N_{CS}(t=0)-N_{CS}(t=\infty)\ee
  is 1/2, since the final state is made of weak fields correponding to $N_{CS}(t=\infty)=0$. 
  But it is not so, and $\Delta N_{CS} <1/2$. Yet the number of fermions pulled out from the Dirac sea is 
  still integer -- one per fermion species -- as explained in detail in~\cite{Shuryak:2002qz} . 
  
\section{The  CP-violating effective actions} 

In the last decade certain efforts were made to calculate the so called 
 CP violating effective actions. 
Technically the effect 
originates from the fermionic loop diagram, representing the fermionic determinant,   
 in some
smooth gauge field background. 

On general ground, the CP odd effects  require at least 4  CKM matrices, so they may in principle appear
starting from the 4-th order in weak interaction.
However, 
according to explicit calculations  \cite{Hernandez:2008db} in this leading order the
result  vanishes.

The first results  in the next-to-leading sixth order, with one extra neutral $Z$ boson vertex, were reported in \cite{Hernandez:2008db}.
The result for the effective action operator involves a dimension-6 operator 

\bear
\label{1}
&&L_{CP}=C_{CP}\epsilon^{\mu\nu\lambda\sigma}\nonumber\\
&&\times\left[ Z_\mu W_{\nu\lambda}^+W_\alpha^-\left(W_\sigma^+W_\alpha^-+W_\alpha^+W_\sigma^-\right)+{\rm c.c.}\right]
 \label{eqn_LCP}  \eear
containing four charged gauge boson fields $W$ fields and one neutral $Z$.
The coefficient has the following form
\be C_{CP} =J {3 \over 2^9 \pi^2}  {\kappa_{CP}(m_q) \over m_c^2} \approx 2.2 \times 10^{-8} GeV^{-2}\ee
where $J=3*10^{-5}$ is the well known Jarlskog invariant combination of sin and cos of the CKM matrix angles.
The coefficient $\kappa_{CP}(m_i)$, normalized at
the charm quark mass  $m_c$   as the  scale of the calculation,  
is a complicated function of  all quark masses $i=s,c,b,t$ given in the Appendix of \cite{Hernandez:2008db}.
It is numerically equal to $   \kappa_{CP}\approx 9.87 $.

Subsequent investigations in \cite{GarciaRecio:2009zp}  have not confirmed a non-zero coefficient
for this operator, but came up instead with a set operators  
of dimension 6 possessing  a completly different structure. 
Another group \cite{Brauner:2012gu} confirmed their finding. Remarkably, all the
13 operators $O_i$ found  are C-odd and P-even while the
above-given (\ref{eqn_LCP}) is P-odd and C-even. 
While we are unaware of a clear explanation for why this is the case, we note that
(\ref{1}) is a  variant of the Wess-Zumino  contribution to the effective action.

Which operators are the correct ones is still to be sorted out. Fortunately, as we found,
their matrix elements are not that different, and the magnitude of the result depends much more
on the coefficients of those operators, to which we turn later.


\section{CP-odd effect in an exploding sphaleron}
\subsection{The matrix elements}

At this point, we are  ready to combine two ingredients of the problem, the
semiclassical sphaleron explosion and the effective next-to-leading  order CP-odd Lagrangians.

Sphalerons -- the tops of the sphaleron path from one valley to another along the topological Chern-Simons coordinate --
are the strongest non-perturbative fluctuations of the gauge field
in the electroweak primordial plasma. So, their field should dominate the dimension-6 operators
of the next-to-leading effective action. 

This turns out to be so for operators $O_i$ suggested in \cite{Brauner:2012gu}, but not 
for the operator  (\ref{eqn_LCP}) which requires a non-zero
electric fields (or, so-to-say, the ``time arrow"). Fortunately, as we have shown above, 
 during the  sphaleron explosion there is a non-zero electric field  present. 
 
\begin{figure}[h]
\begin{center}
\includegraphics[width=8cm]{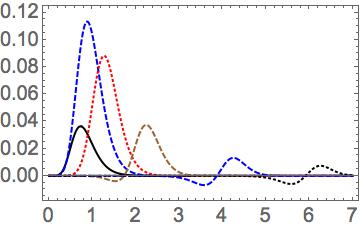}
\caption{The radial dependence of the $r^2 L_{CP}$, the CP-odd Lagrangian (\ref{1}) without the coefficient, at various time moments
$t=0.1,0.5,1.,2.,4.,6$ (in units of the sphaleron size parameter $\rho$), shown respectively by  
(thick black solid), (blue dashed), (red dotted), (brown
dashed), (blue dashed) and (black dotted) curves.}
\label{fig_r2_CP}
\end{center}
\end{figure}

Substituting  our semiclassical exploding sphaleron solution (\ref{eqn_field})
into the CP-odd operators results in a quite lengthy analytic expression, not to be given here.
One example -- for the operator  (\ref{eqn_LCP}) -- is
 plotted as a function of $r$
 at several time moments in Fig.~\ref{fig_r2_CP}. According to the ``time arrow" argument,
 at the initial ``sphaleron time" $t=0$ it vanishes. However, it quickly grows from it, reaching a certain maximal value.
 At large time the fields magnitude naturally decreases, but when plotting $r^2 L_{CP} $ as we do, including 
 the volume of the spherical shells,  one finds that this decrease is indeed compensated. The shape at large time
 is stabilized into some sign-changing pattern, roughly corresponding to that of the  electric field.

The exploding sphaleron solution we describe is defined in the infinite volume of the symmetric phase
(in fact we not only ignored the VEV of the Higgs field, but its presence altogether). In the application we 
intend to use it -- the ``hot spots" during the cosmological out-of-equilibrium transition -- the space and time 
are limited by their walls and lifetime. For definiteness, 
for the CP-violating action we will mention the value of the integral inside the domain
$ 0<r,t< 3\rho$
\be A_{3,3}=\int_{0<r,t< 3\rho} dt dr 4\pi r^2 L_{CP} \approx {1.54 \over  \rho^2} \label{eqn_act} \ee
which turned out to 
 depend rather weakly on the modified size and shape of the integration domain.
 Using this value and the coefficient given above, putting the sphaleron scale
 $\rho=1/m_c$, one finds the CP violating action to be \be \delta_{CP}  \sim 10^{-7} \ee
 
The other proposed operators  produce similar integrated results for the matrix elements,
 and since all of them have  dimension 6, these integrals all scale as $\rho^{-2}$.

\subsection{Scale dependence of the coefficients }

The original papers which made generic estimates of CP violation at the electroweak scale
obtain very small values, precluding its use as a possible explanation of the baryon asymmetry 
of the Universe. Yet Smit \cite{Smit} suggested  that  the argument  is not so simple, since the effect 
of CP violation is scale dependent, increasing substantially at lower scales to reach the maximum value
of $\delta_{CP} \sim J\sim 10^{-5}$.

The operators $O_i$ contributing to the effective Lagrangian  are multiplied by Wilsonian  coefficients
$c_i(\mu)$ which include the contributions from all  scales above the ``normalization scale" $\mu$.
While each coefficient is a slightly different function of $\mu$, there are generic features common to them all,
which we try to derive below.

Suppose one is interested in  CP violation in a specific background gauge field composed of 
the weak fields $W_\mu(x)$ and $Z_\mu(x)$
(with perhaps also known  Higgs $\phi(x)$, if needed). In general, 
the determinant of the Dirac operator  can be 
written in a box structure in the left-right spinor notations
\be det\left(  \begin{array}{cc} 
i\Dslash & M \\
M^+ & i\partslash \\
\end{array}
\right) 
= det( i\partslash) \,det\left(i\Dslash+M {1 \over  i\partslash}M^+ \right)
\ee
where $M$ is a mass matrix in flavor space and  the slash here and below means the convolution with the Dirac matrices,
e.g. $\Dslash=D_\mu \gamma_\mu$. 

The long covariant derivative 
involves the weak gauge fields. 
Let us use a representation in which this operator is diagonalized
\be i\Dslash  \psi_\lambda(x)=\lambda  \psi_\lambda(x) \ee
Its two sub-operators, $\partslash$ and $ \Wslash$ are not in general 
diagonal in this basis, but for our qualitative argument we will only include their
diagonal parts
\be  <\lambda |i \partslash | \lambda' > \approx \pslash \, \delta_{\lambda \lambda'}, \,\,\,\,
<\lambda | \Wslash | \lambda' > \approx \xi \lambda\, \delta_{\lambda \lambda'} \ee
where $\pslash,\xi$ are in general some functions of $\lambda$, ignoring the
non-diagonal ones . In this approximation 
the corresponding  (Eucidean) propagator  describing a quark of flavor $f$ propagating  in the background can be represented as the usual
 sum over modes 
\be S(x,y) \approx  \sum_\lambda {\psi^*_\lambda(y) \psi_\lambda(x) \over \lambda +M \pslash^{-1} M^+} \ee
where the right-handed operator $i\partslash$ 
is approximated by its diagonal matrix element in the $\lambda$-basis. Throughout, we will trade
the geometric mean appearing in all expressions
$\sqrt{\pslash \lambda} \rightarrow \lambda$, for simplicity.

The generic fourth-order diagram in the the weak interactions, contain at least  four CKM matrices, 
and takes in the coordinate representation the form
\begin{eqnarray}
\label{box}
 &&\int (\Pi d^4 x_i)  \, Tr[ \Wslash(x_1) \hat V  \hat S_u(x_1,x_2)  \Wslash(x_2)  \nonumber \\
&&\hat V^+  \hat S_d (x_2,x_3)\Wslash(x_3)  \hat V  \hat S_u  (x_3,x_4)\Wslash(x_4)   \hat V^+  \hat S_d(x_4,x_1)] \nonumber
\end{eqnarray}
Here $V$ is the CKM matrix, with the hats and the propagators
labels $u,d$  indicating that they are $3\times 3$ matrices in flavor subspace.
The trace is over both spin and flavor indices.
If one considers the  next order diagrams, with $Z,\phi$ field vertices, the expressions are generalized straightforwardly. 

The spin-Lorentz structure of the resulting effective action is very complicated. However, to understand the
scale dependence we will make  a second strong simplifying assumption.
With this in mind, one can use the orthogonality condition of the different $\lambda$-modes 
and perform the integration over coordinates, producing  a simple expression, with a single sum over eigenvalues
$ \sum_\lambda F(\lambda)$ with the following generic function
\be
F(\lambda) =
\lambda^4 Tr\left(  \hat V S_d  \hat V^+ S_u  \hat V S_d  \hat V^+ S_d \right)
\ee
This is the box diagram associated to (\ref{box}) in the $\lambda$-representation, which generalizes the momentum representation valid
only for  constant fields. Unlike momenta, the spectrum of the Dirac eigenvalues $\lambda$ may have various
spectral densities $d(\lambda)$. In particular, there is a zero mode, corresponding to  the zero mode in the original 4-dimensional
symmetric case,  describing the  fermion production.

Whatever  the physical meaning of the spectral density
$d(\lambda)$, the point is that one can perform the multiplication of the flavor matrices
and extract a universal function of $\lambda$, that describes the dependence of the  
CP violation contribution on the underlying scale.
Using the standard form of the CKM matrix $\hat V$, in terms of the
known three angles and the CP-violating phase $\delta$, and also the six known quark masses, 
one can perform the multiplication of these 8 flavor matrices and identify the lowest order  CP-violating term of the result.
Performing the multiplication in the combination above one finds a
complicated expression which does $not$ have $O(\delta)$ term,
so there is no lowest order CP violation. This agrees with a statement from
\cite{Smit} that the leading fourth order diagram generates no operators in the effective action.

Higher order diagrams however all have such contributions. We generated a number of those: the simplest
turned out to be the sixth-order diagram with four $W$ vertices and two $Z$, namely
%
\be
F_{ZZ}(\lambda) = 
\lambda^6 Tr\left(  \hat V S_d  \hat V^+ S_u  \hat V S_d Z S_d  \hat V^+ S_u Z S_u \right)
\ee
Now the flavor trace has the lowest order CP violation 
described by the following symmetric expression
\begin{widetext}
\begin{eqnarray}
\label{FZZ}
 Im F_{ZZ}(\lambda) =  2  \lambda^6 {J (m_b^2 - m_d^2) (m_b^2 - m_s^2) (m_d^2 - m_s^2) (m_c^2 - m_t^2) (m_c^2 - m_u^2) (m_t^2 - m_u^2) \over \Pi_{f=1..6} (\lambda^2 +m_f^2)^2 } 
 \end{eqnarray}
The numerator is the familiar Jarlskog
combination of the CKM angles and differences of masses squared,
appended by a nice symmetric denominator.
(As expected, the effect vanishes when the mass spectrum 
 of either u-type or d-type quarks gets degenerate.)
 
 If instead of two $Z$ vertices there are two $\phi$ ones, coupled to
 fermions in proportion to their masses, the flavor trace is
 \be
F_{\phi\phi}(\lambda) = 
\lambda^6 Tr\left(  \hat V S_d  \hat V^+ S_u  \hat V S_d m_d S_d  \hat V^+ S_u m_u S_u \right)
\ee
 resulting in a more complicated expression 
 \be Im F_{\phi\phi}=- J {  \Delta_1 \Delta_2 \over \Pi_{f=1..6} (\lambda^2 +m_f^2)^2 }  \ee
$$ \Delta_1= (m_b - m_d) (m_b - m_s) (m_d - m_s) (m_c - m_t) (m_c - m_u) (m_t - 
      m_u) $$
      $$
  \Delta_2=    m_b^2 (m_d m_s - \lambda^2) - \lambda^2 (m_d^2 + m_d m_s + 
         m_s^2 + \lambda^2) + 
      m_b (m_d^2 m_s - m_s \lambda^2 + $$
  $$ +       m_d (m_s^2 - \lambda^2))) (m_c^2 (m_t m_u - \lambda^2) - 
\lambda^2 (m_t^2 + m_t m_u + m_u^2 + \lambda^2) + 
      m_c (m_t^2 m_u - m_u \lambda^2 + m_t (m_u^2 - \lambda^2)) 
  $$  
which  has similar symmetries and possesses the same denominator, mostly responsible for its scale 
dependence.
 \end{widetext}

\begin{figure}[h!]
\begin{center}
\includegraphics[width=7.cm]{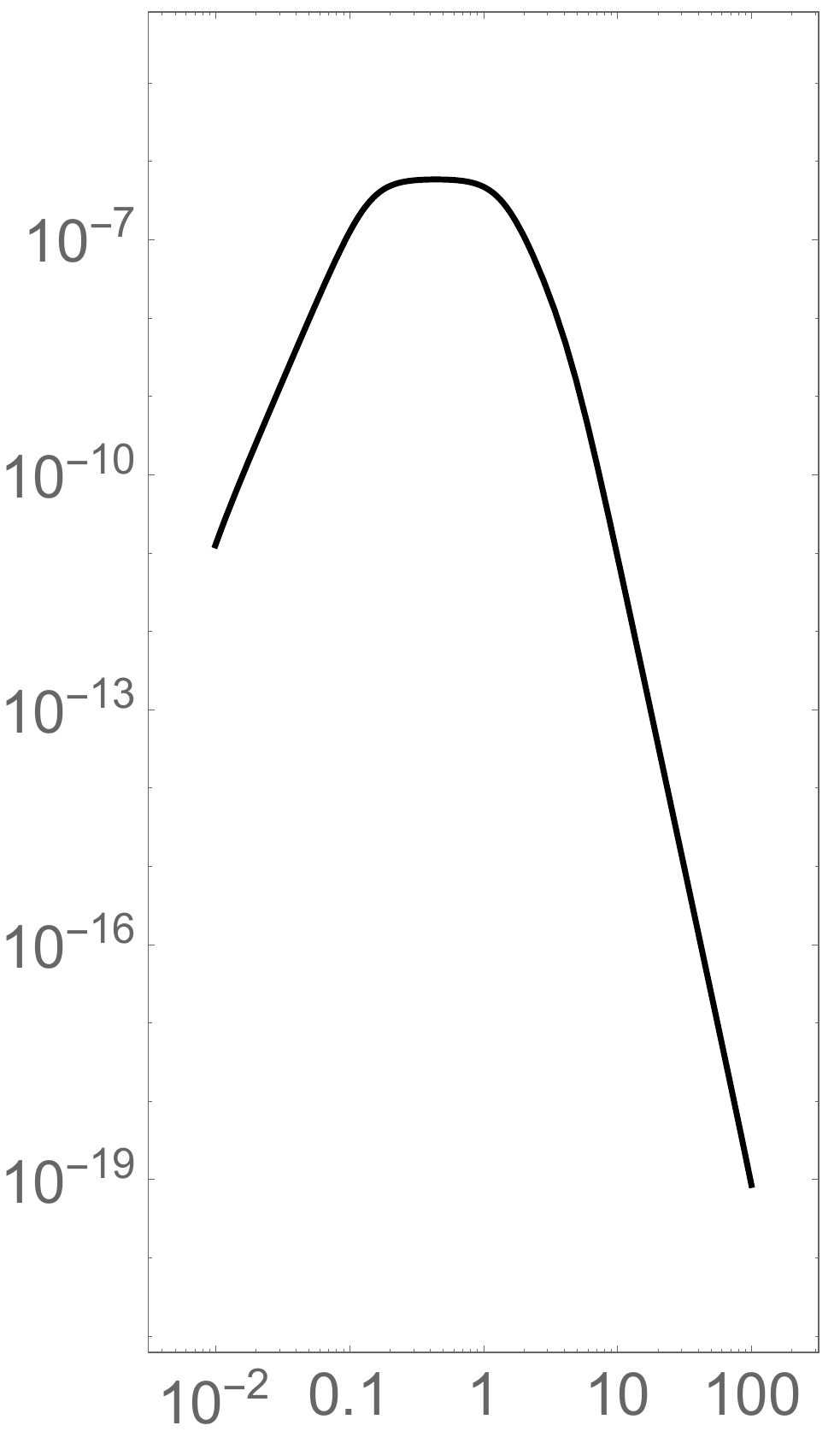}
\caption{The CP-violating part of the $W^4 Z^2$ diagram $ Im F_{ZZ}(\lambda)$ versus $\lambda$ (GeV).}
\label{fig_Fcp_lambda}
\end{center}
\end{figure}

A plot of (\ref{FZZ})  is shown in Fig.\ref{fig_Fcp_lambda}. Because of the cancellation
between different quark flavors, it is very small at large $\lambda$, about $10^{-19}$ at the electroweak scale of
$\lambda\sim 100 \, GeV$ which corresponds to the right-hand-side of the plot, which can be called ``the Jarlskog regime".
At the scale near and below $\lambda =1\, GeV$ one finds an alternative ``Smit regime", in which the CP violation is twelve orders of magnitude larger!

All the quark masses are taken at their physical values in our world, that is 
for a Higgs field equal $v$. However,  in a ``hot spot" in which the sphaleron transition happens,
the Higgs expectation value is smaller than in the broken phase, $\phi< v$, by  a factor. One can take care 
of this through the rescaling $\lambda\rightarrow \tilde\lambda=\lambda/(\phi/v)$. Since the function is dimensionless,
its values are preserved, and  the plot just moves horizontally as a whole by this factor (not shown on the plot). 

  One thing to check is whether the calculation for the operator (\ref{eqn_LCP}) done above 
  agrees with this curve or not. The calculation was done at a normalization point
equal to the  charm quark mass $m_c\approx 1.4\, GeV$, and the result was 
  $\delta_{CP}\approx 10^{-7} $. This is  approximately  in agreement with the scale dependence
  shown in Fig.~\ref{fig_Fcp_lambda}.

The temperature dependence of the CP violation coefficients
should also follow from  this universal function. Indeed,  if one uses for a fermion  the lowest Matsubara frequency $\lambda=\pi T$, our generic function is indeed similar  to the operator coefficients $c_i(T)$ of the sixth-order terms calculated in 
\cite{Brauner:2012gu}. 

Of course, different background  fields have different spectral density of states $dN/d\lambda=d(\lambda)$, and convolution
of that with our generic function $F(\lambda)$ is somewhat different. However, for (non-topological) background fields one expect the density of states to be a smooth function. (For example if the fields are constant and the eigenvalues are 4-momenta,
then  $d(\lambda)\sim \lambda^3 $,
and this behavior is expected at large values for any smooth fields.) A convolution of such a density with  the very rapidly
changing generic funciton $F(\lambda)$ will only modify the numbers mentioned above slightly.

This  dependence on the scale displays a generic phenomenon of the coefficient functions on the scale.

\section{The zero modes, fermion production and CP violation on the in/out-going lines}

We stated at the end of the previous section that for non-topological background fields
the density of the eigenvalues of the Dirac operator $\Dslash$ are smooth. 
In this section we focus on the topologically nontrivial backgrounds -- of which the exploding sphaleron
is one -- which, as demanded by index theorems, have certain numbers of fermionic zero modes, 
states with $\lambda=0$.

The naive substitution of $\lambda=0$ produces zero, but this is misleading.
As shown by 't Hooft in 1970's, in this case the expressions need special care.
The physics is that of  fermion production, and the zero mode describes the state in which
that happens. Formally it means that the fermionic loop gets ``opened" into a line. One end of it  corresponds to
 the initial fermionic state at $t=0$, and the other end  corresponds to  its asymptotic in/outgoing states at $t\rightarrow \mp \infty$,
 when the background field becomes negligibly weak.

In the case of the exploding COS sphaleron this zero mode was explicitly constructed in our 
work  \cite{Shuryak:2002qz}. It starts from the zero energy state bound to the static sphaleron at $t=0$ 
and turn to a 
positive energy fermions flying away at late time. The momentum spectrum
is just given by the (``amputated") Fourier transform of the mode at late time
\be {dN \over d^3p} \sim \left| \pslash \psi_0(p, t \rightarrow \infty)  \right|^2 \ee
There are 9 quarks and 3 leptons which are produced in this way.
Of course the expression above is written in the lowest order approximation in which 
all fermionic masses, CKM matrices etc are all neglected.

The CP violation which happens on these incoming/outgoing lines, appears only when 
the corrections to this semiclassical expression are included, starting from fourth-order.
The way we will take those into account closely follow what was done in \cite{Burnier:2011wx},
in which there were no sphalerons, but a decaying top
 quarks escaping the ``hot spots" into the ambient plasma. 
 
\begin{figure}[h]
\begin{center}
\includegraphics[width=8cm]{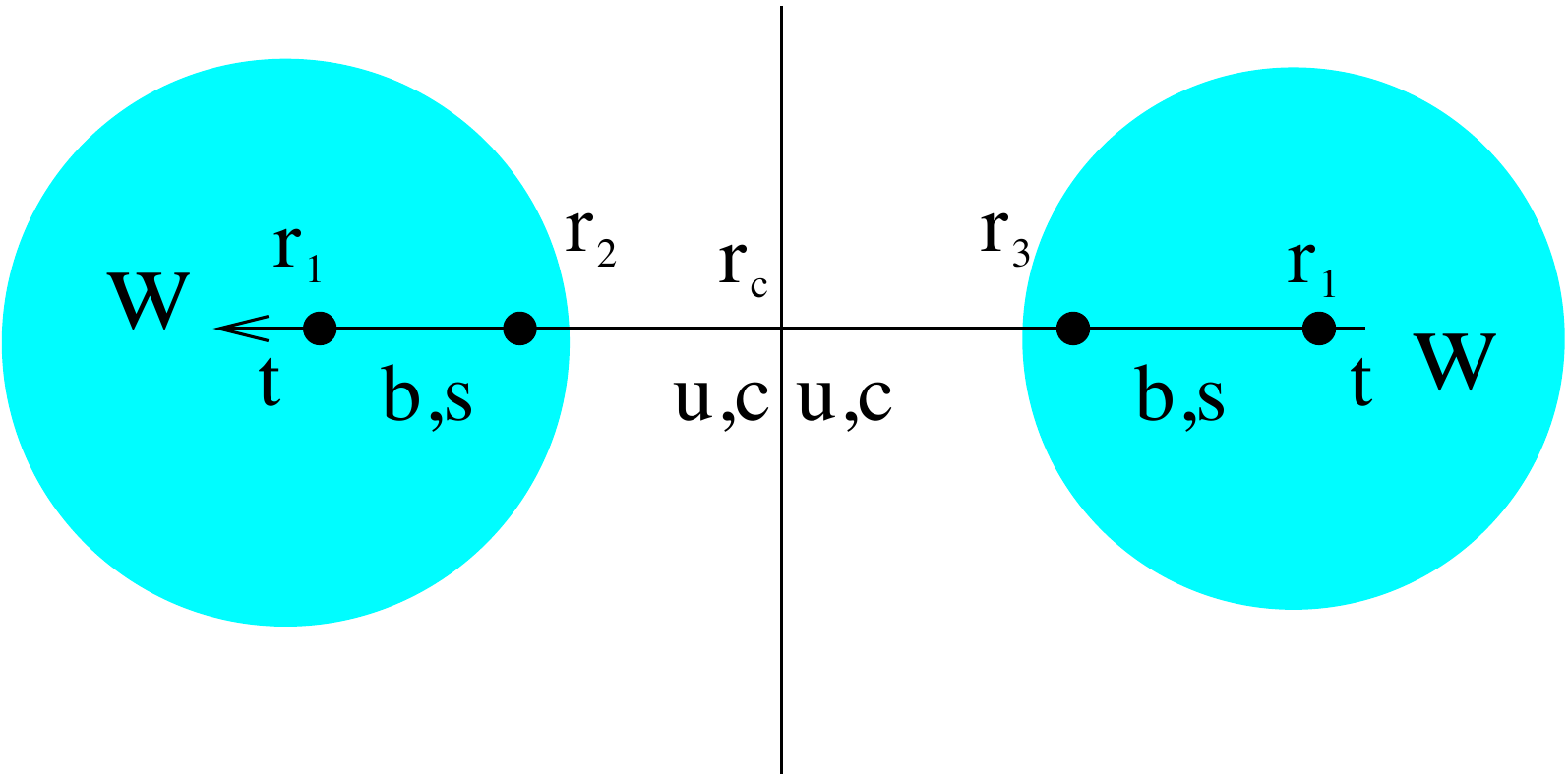}
\caption{Schematic shape of the fourth-order process involving only quarks of the 2nd and 3rd generations. The blue shaded regions on the left  and right represent region of strong background gauge fields (indicated by W in the figure) of the exploding
sphaleron. The vertical line is the unitarity cut. Four black dots indicate 4 points,  $r_i,i=1..4$, where 
quarks undergo additional weak  interactions with the background. It is assumed that the in/out going quarks are tops,
and the intermediate ones are either $b,s$ or $c,u$. (the small contribution from  $d$ is neglected).}
\label{fig_2bags}
\end{center}
\end{figure}

 Since we are not aiming for a quantitative calculation but rather an estimate,
 we will not discuss all possible diagrams with 4 weak interactions, but focus on one of them
depicted in Fig.~\ref{fig_2bags}, in which 4 final state quark rescattering
are distributed into two pairs, in the amplitude and the conjugated amplitude.
The intermediate properties --
the flavors of the quarks in between, locations of the points $r_i$ etc -- can be summed over freely,
but  quantum numbers of the particles going through the unitarity cut in the middle should 
of course be carried continuously from one side to the other. This allows us to connect both amplitudes to a single
matrix product, beginning and ending from the fermionic zero mode
 \footnote{The propagation through the cut in Fig.~\ref{fig_2bags} occurs on mass-shell with the PP-part  in the propagator
 excluded. However, for the estimate to follow this constaint is not important.}

\begin{widetext}
\be P_t=\int A^+_t A_t = \psi_{0t}^+(r_1)  \Wslash^-(r_1) \hat V^+ S_d(r_1,r_2) \Wslash^+(r_2) \hat V S_u(r_2,r_3)
  \Wslash^-(r_3) \hat V^+ S_d(r_3,r_4)   \Wslash^+(r_4) \hat V  \psi_{0t}(r_4) \ee
\end{widetext}
The subscript $t$ indicates that the in/out quark is the top quark.
The summation over the flavor indices of the 4 CKM matrices is assumed.  
 Since there is a CP-violating phase $\delta$, this probability for a top quark on a sphaleron
 is different from that of the anti-top on an anti-sphaleron
 (not written explicitly), and therefore the probabilities to increase or decrease the baryon number  
 are slightly different.
 
 The difference between them
 $P_t -P_{\bar{t}}$ depend on the fact that the quark masses in the propagators connecting
 the points   are  different. In the coordinate representation we now use,  this comes from
 the phase factors which can be semiclassically evaluated as
 \begin{widetext}
   \be
  S_{12}=exp\left(i \int_{r_1}^{r_2}p(x)dx\right)\approx exp\left(i \int_{r_1}^{r_2}\left(E-{\frac{m_i^2(x) }{ 2E}}\right) dx\right)
  \ee
  \end{widetext}
  where $E$ is the quark energy, and the approximation implies that all lower quark flavors are  light enough
  compared to $E$. Technically, the eikonal propagation is on the light cone, but for our simplifying estimates the difference
  is not consequential.
  
  The flavor structure of the $P_t -P_{\bar{t}}$ comes from the 4 CKM matrices and the propagators, which
  was shown in \cite{Burnier:2011wx} to reduce to the following expression
    \bear
  M_t- M_{\bar{t}}= 2 i J (S^u_{23}-S^c_{23}) \nonumber \\ 
 (-S^s_{34} *S^b_{12}-S^d_{34} *S^s_{12}+S^d_{34} *S^b_{12}   \nonumber \\ 
 +S^d_{12} *S^s_{34}-S^d_{12} *S^b_{34}+S^s_{12} *S^b_{34})
 \label{eqn_Mt-Mantit}
   \eear
  where $J$ is once again the Jarlskog factor. 
 we note further, that if the $u,c$ quarks would have the same mass, 
the first bracket vanishes. This is in agreement with general arguments that 
any degenerate quarks should always nullify the CP-odd effects,
as the CP odd phase can be rotated away already in the CKM matrix itself.

The last bracket in (\ref{eqn_Mt-Mantit}) contains interferences of different down quark species.
We note that there are 6 terms, 3 with plus and 3 with minus.
 Each propagator, as already noticed in the preceeding section, has only small corrections coming from the quark masses. Large terms which are flavor-independent
  always cancel out, in both brackets in the expression above.   Let us look at only the terms which contain the heaviest $b$ quark in the last bracket, using the propagators in the form 
  
\be S^q_{ij}=exp\left( \pm i \delta_{ij}  \right)=  exp\left( \pm i \frac{m_b^2}{2 E} r_{ij} \right), \ee
  where $\pm$ refers to different signs in the amplitude and conjugated amplitude and $r_{ij}=r_j-r_i$. Note that the sign of the phase between points $r_2$ and $r_3$ can be positive or negative as it results from a subtraction of the positive phase from $r_3$ to the cut $r_c$ with the negative phase from the cut $r_c$ to $r_2$. The terms containing odd powers  in $r_{23}$ therefore 
  should vanish in the integral, and the lowest contribution is quadratic.
  Considering all phases  to be small due to $1/E$  and using the mass hierarchy $m_b \gg m_s \gg m_d$,  
  we pick  the leading contribution of the last bracket in (\ref{eqn_Mt-Mantit}) which has $r_{23}^2$
  the 4-th power in the last bracket,  and the 6-th order in the phase shift in total
  \be 
   P_t- P_{\bar{t}}\sim  J   \frac{ m_b^4 m_c^4 m_s^2 r_{23}^2 r_{12}  r_{34} (r_{12}+   r_{34})}{ E^5}.
\label{F}
  \ee 
 Note that all distances in this expression are defined to be positive and the sign in the last bracket is plus, so unlike all the previous orders in the phase expansion, at this order we have a sign-definite answer with no more cancellations possible.
 
 This expression can now be averaged over the position of the 4 points
\begin{widetext}
 \be 
  P_t- P_{\bar{t}}\sim  \int (\Pi_{i=1}^4 d^4 r_i )    r_{23}^2 r_{12}  r_{34} (r_{12}+   r_{34})  \psi_{0}^+(r_1)  \Wslash^-(r_1) S_0(r_1,r_2) \Wslash^+(r_2)  S_0(r_2,r_3)  \Wslash^-(r_3)  S_0(r_3,r_4)   \Wslash^+(r_4) 
 \psi_{0}(r_4)
\ee \end{widetext}
in which we omitted $J$ and the mass factors, already given in the previous expression. Zero near the propagators
indicate that all of those are considered to be massless, as the mass-dependent phases were already
included above.  In the coordinate representation used here, they are simply $S_0(r)=(r_\mu \gamma_\mu)/(2\pi^2 r^4)$.  
The remaining indices are the spinor ones, and the trace is elementary. The multidimensional
point integrals can be done numerically: all $r_i$ have the scale given by the sphaleron size parameter $\rho$.

\section{CP violation in the cold electroweak cosmology}

So far we have discussed a general academic case, with the exploding sphaleron background having an  unspecified scale.
 Now we take this scale to be as given in 
 the numerical simulations 
\cite{GarciaBellido:1999sv}  for the cold electroweak transition. Mapping of these sphalerons  to the
analytic COS solution  \cite{Flambaum:2010fp} resulted in the
 mean sphaleron parameter $\rho$ being
\be \rho \approx {3.9 \over m} \approx {1 \over 63 \, GeV} \ee
where  $m\approx v \approx 246\, GeV$ is the ``weak unit" used in the simulations.  
For definiteness, let us associate  the typical eigenvalue $\lambda$ with the square root of the magnetic field
of the sphaleron
\be \lambda \approx \sqrt{B}\rightarrow {130 \, GeV \over 1+r^2/\rho^2} \label{eqn_lambda_r} \ee

Let us start our discussion from the contribution of the non-zero modes. 
It is clear that at the sphaleron center, $r=0$,  the scale $\lambda\approx 130 \, GeV$ is way too high:
according to Fig.~\ref{fig_Fcp_lambda}  the CP violation at this scale is  $10^{-18}$ or so.
Stronger CP violation can perhaps be produced either (i) at the sphaleron's periphery, where the field is weaker;
or (ii) using the fact that sphalerons fluctuate and perhaps there is some tail of the distribution over $\rho$
that extends to smaller values.
Let us discuss them subsequently.

As already emphasized above, all sphaleron transitions observed in numerical simulations happen inside
the so called ``hot spots". Their size is larger than the size $\rho$ of the sphalerons, but only by about a factor 2 or so. 
Substituting this into (\ref{eqn_lambda_r}) one finds that the smallest $\lambda$ available at the sphaleron periphery
is of the scale $\lambda \sim 30\, GeV$ or so. The corresponding magnitude of the CP violation deduced from
the plot in Fig.~\ref{fig_Fcp_lambda} is  about $ 10^{-16}$, again too small.

Outside of the ``hot spots" there is what is called the ``ambient plasma". When fully equilibrated with all fermionic 
and gluonic degrees of freedom, the plasma total energy is black-body with the  temperature
$T\sim 35\, GeV$. Quarks moving in the plasma get rescattered, with the largest effect
coming from the gluons and inducing the so-called Klimov-Weldon mass 
\be M_q^2= {g_{strong}^2 T^2 \over 6} \approx  (14\, GeV)^2 \label{thermal_mass}\ee
If this mass provides a  cutoff for the eigenvalue spectrum with  $ \lambda > M_q $
(as it does for quark energies), then the plot in Fig.~\ref{fig_Fcp_lambda} 
suggest a CP violation effect of about $10^{-13}$.  


Now, what would happen if the sphaleron sizes, the field magnitude or the eigenvalues cutoff  can fluctuate significantly,
reaching values much smaller than the average?
Inverting the logic, one can start with
 the generic  plot Fig.~\ref{fig_Fcp_lambda} and ask, what is the scale corresponding to  a
``minimally sufficient CP violation" contribution,   say $10^{-8}$? The answer is 
 $\lambda \sim 5\, GeV$. A probability of  fluctuations in which the Dirac operator has  eigenvalues
in  this scale  range,  has yet to be studied in numerical simulations.

Now we turn to the CP violation from the zero mode sector, the final state interaction accompanying a fermion production.
The estimates (ignoring numerical factors~\footnote{Let us comment on the factors of $\pi$.
The four integrals over $d^4 r$ will bring factors of $(2\pi^2)^4$ from solid angles, but three of these factors are cancelled by
similar factors in the propagators, and the last by the normalization of the zero mode.  
}) for it derived above reads
\be \delta_{CP}\sim { J m_b^4 m_c^4 m_s^2 \rho^5 \over E^5} \ee
The minimal 
value of the energy of the propagating quark $E$ should  be the  thermal fermion mass
in the ambient plasma (\ref{thermal_mass}). 
Yet even if we go to lower scale,
 $E_{min} \sim 5\, GeV$,  the result  from the expressions above is however still too small
$ \delta_{CP}\sim 10^{-16} $,
way below the values needed for
an explanation of the baryon asymmetry.

Summarizing this paper we conclude that CP violation near sphaleron transitions, with 
the average parameters suggested by numerical simulations, are too small to explain the observed baryon asymmetry.
However, the effect of a sufficient magnitude can be found, provided there is a sufficient 
spectral strength of the Dirac operator at a scale of about $5\,GeV$.

 {\bf Acknowledgements.}
 ES thanks Michael Ramsey-Musolf and all other organizers of the Baryogenesis Program at MIAPP Munich, June 2016,
   for the invitation to the program, which prompted
 us to think about this problem again. 
This work is supported in part by the U.S. D.O.E. Office of Science,  under Contract No. DE-FG-88ER40388.


\end{document}